\documentclass[12pt]{article}
 \pdfoutput=1
\usepackage{amssymb}
\usepackage{amsmath}
\usepackage{amstext}
\usepackage{graphicx,epsfig}
\usepackage{epsfig}
\usepackage{verbatim} 
\usepackage{fancybox}
\usepackage{color}
\usepackage{ulem}
\usepackage{enumitem}
\usepackage{subfigure}
\usepackage{bbm}
\usepackage{parskip}
\usepackage{cite}

\newcommand{\Comment}[1]{{}}
\definecolor{MyDarkBlue}{rgb}{0.15,0.15,0.45}
\usepackage[linktocpage=true]{hyperref}
\hypersetup{
colorlinks=true,
citecolor=MyDarkBlue,
linkcolor=MyDarkBlue,
urlcolor=MyDarkBlue,
pdftitle={Amplitudes and 4D Gauss--Bonnet Theory},
pdfsubject={hep-th}
}

\newcommand{\be}{\begin{equation}}
\newcommand{\ee}{\end{equation}}
\newcommand{\bea}{\begin{eqnarray}}
\newcommand{\eea}{\end{eqnarray}}
\newcommand{\beas}{\begin{eqnarray*}}
\newcommand{\eeas}{\end{eqnarray*}}
\newcommand{\nn}{\nonumber}

\def\({\left(}
\def\){\right)}

\newcommand{\pb}{\mathbf{p}}

\numberwithin{equation}{section}

\begin{document}

\begin{center}
{\LARGE \bf{Amplitudes and 4D Gauss--Bonnet Theory}}
\end{center} 

\vspace{2truecm}

\thispagestyle{empty}
\centerline{{\Large James Bonifacio,${}^{\rm }$\footnote{\href{mailto:james.bonifacio@case.edu}{\texttt{james.bonifacio@case.edu}}} Kurt Hinterbichler,${}^{\rm }$\footnote{\href{mailto:kurt.hinterbichler@case.edu}{\texttt{kurt.hinterbichler@case.edu}}} Laura A. Johnson${}^{\rm }$\footnote{\href{mailto:laura.johnson2@case.edu}{\texttt{laura.johnson2@case.edu}}}}}
\vspace{.7cm}

 \centerline{{\it ${}^{\rm }$CERCA, Department of Physics, Case Western Reserve University,}}
 \centerline{{\it  10900 Euclid Ave, Cleveland, OH 44106}} 
 \vspace{.35cm}

\begin{abstract}

It has recently been argued that there may be a nontrivial four-dimensional limit of the higher-dimensional Gauss--Bonnet and Lovelock interactions and that this might provide a loophole allowing for new four-dimensional gravitational theories, possibly without a standard Lagrangian.  We investigate this claim by studying tree-level graviton scattering amplitudes, allowing us to draw conclusions independently of the Lagrangian.   By taking four-dimensional limits of higher-dimensional scattering amplitudes of the Gauss--Bonnet theory, we find four-dimensional amplitudes that are different from general relativity; however, these amplitudes are not new since they all come from certain scalar-tensor theories.  The nontrivial limit that does not lead to infinite strong coupling around flat space leads to $(\partial\phi)^4$ theory.
We argue that there cannot be any six-derivative purely gravitational four-point amplitudes in any dimension other than those coming from Lovelock theory by directly constructing the on-shell amplitudes.  In particular, there can be no new such amplitudes in four dimensions beyond those of general relativity. We also present some new results on the spin-averaged cross section for graviton-graviton scattering in general relativity and Gauss--Bonnet theory in arbitrary dimensions.   

\end{abstract}

\newpage

\section{Introduction}
\parskip=5pt
\normalsize

According to Lovelock's theorem \cite{Lovelock:1971yv,Lovelock:1972vz}, the most general theory of a single interacting massless graviton in any dimension leading to second-order equations of motion is given by a sum of the Lovelock--Lanczos terms,
\be   {\cal L}^{(n)}={n!\over 2^{n/2}} \delta^{[\mu_1}_{\nu_1} \cdots \delta^{\mu_n]}_{\nu_n}R_{\mu_1\mu_2}^{\ \ \   \ \ \nu_1\nu_2}R_{\mu_3\mu_4}^{\ \ \ \ \ \nu_3\nu_4}\cdots R_{\mu_{n-1}\mu_n}^{\ \ \ \ \ \ \ \nu_{n-1}\nu_n},\ \ \ n=0,2,4,\ldots \,.
 \label{lovelocklag} \ee
In $D$ dimensions, the terms with $n\geq D$ are total derivatives that do not contribute to the equations of motion or on-shell amplitudes.  
In four dimensions, only ${\cal L}^{(0)}=1$ and ${\cal L}^{(2)}=R$ are nontrivial, and the theory reduces to general relativity (GR) with a cosmological constant, and thus GR is supposed to be the unique ghost-free theory of a single interacting  massless spin-2 particle in four spacetime dimensions.

Since the higher-order Lovelock terms are trivial when $D=4$, observables calculated in $D$ dimensions including these terms should lose their dependence on them as $D\rightarrow 4$.   Recently there has been renewed interest, sparked by Ref.~\cite{Glavan:2019inb}, of the possibility of scaling the coefficients of these higher-order terms by factors of $D-4$ in such a way that there is a nontrivial finite remainder when $D=4$ \cite{Tomozawa:2011gp,Cognola:2013fva,Glavan:2019inb}.  For example, taking the simplest case of Gauss--Bonnet theory, where we only add the quadratic in curvature Lovelock term (we will stick to flat space and ignore the cosmological constant),
\be S={M_P^{D-2}\over 2}\int d^Dx\sqrt{-g} \left(R+\alpha\, {\cal G}\right),\ \ \ \  {\cal G}\equiv {\cal L}^{(4)}= R_{\mu\nu\alpha\beta}^2-4R_{\mu\nu}^2+R^2,\label{higherda}\ee
one can compute spherical and FLRW solutions in $D$ dimensions and find that the $\alpha$-dependent parts scale like $D-4$.  By replacing
\be \alpha \rightarrow {\tilde\alpha\over D-4}\label{replaceorige}
\ee
and keeping $\tilde{\alpha}$ fixed, one finds finite $\tilde\alpha$-dependent corrections to the GR solutions at $D=4$.  Making the replacement \eqref{replaceorige} directly in the action \eqref{higherda} does not yield a sensible action when $D=4$ because of the singular coupling.

It has been argued that there are no true four-dimensional purely gravitational local equations of motion that could result from such a limit \cite{Gurses:2020ofy,Ai:2020peo}, which would also mean that there is no such four-dimensional Lagrangian.  However, even if such a Lagrangian or equations of motion do not exist, it does not necessarily mean that an interesting theory does not exist.  It could be that the above prescription allows one to compute observables that correspond to some new theory without a standard Lagrangian.  This would be very interesting, because it would explain how such possibilities avoid Lovelock's theorem and were previously missed, and would open the door to many exciting new possibilities.  In fact, as discussed in Refs.~\cite{Casalino:2020kbt,Ai:2020peo}, there would be an infinity of new possibilities because one could take into account any of the higher-dimensional Lovelock terms before taking a $D\rightarrow 4$ limit.  The same possibility could also be present not just for gravitational theories, but for field theories with topological terms that vanish below some critical dimension, such as theta terms, Wess--Zumino terms, or the galileons \cite{Nicolis:2008in, Goon:2012dy} and their multi-field generalizations \cite{Deffayet:2010zh,Hinterbichler:2010xn,Trodden:2011xh}.

Thus we will take the view that there need not be a Lagrangian or equations of motion defined directly in four dimensions, but  we instead use the above prescription for computing observables.  It seems reasonable to require that this prescription apply in complete generality and allow us to calculate anything that we could calculate with a four-dimensional Lagrangian.  

So far this prescription has been studied mostly at the level of highly symmetric solutions and the linear fluctuations around them (see, e.g., Refs.~\cite{Konoplya:2020qqh, Fernandes:2020rpa, Ghosh:2020vpc,Wei:2020ght} and other papers referencing \cite{Glavan:2019inb}).  Here we will go beyond this and study tree-level scattering amplitudes on flat space.  This is equivalent to studying higher-order fluctuations about flat space.  

Amplitudes are basic observables in a gravitational theory defined in flat space, and perhaps the only true observables.
Any prescription for defining a theory with a Minkowski vacuum should thus be able to produce amplitudes that satisfy the usual requirements such as Lorentz invariance, locality, and unitarity.
By studying amplitudes, we also have the advantage of not relying on the existence of a Lagrangian description, and can explore whether there are new non-Lagrangian possibilities in $D=4$ from this prescription.

We start by computing $D$-dimensional amplitudes in the Gauss--Bonnet theory and attempting to take $D=4$ at the end of the calculation to find new four-dimensional amplitudes.  Choices must be made about what $D$-dimensional polarizations and kinematics to use, another sign that the $D\rightarrow 4$ limit is not uniquely defined.  One natural choice is to use $D$-dimensional polarizations that look like ordinary gravitons traveling only in four of the dimensions, with no dependence on the extra dimensions.  We will find that this can give finite $\alpha$-dependent corrections to the amplitude, but it requires a different scaling with $\alpha$ than that required for the black hole solutions.  The resulting amplitude has a pole corresponding to scalar exchange, so it is not a purely gravitational amplitude, and there is in fact a scalar field in the theory.  Other choices of the polarizations give amplitudes involving this scalar as an external state.  
Thus we do not get any new gravitational theory, we get a certain scalar-tensor theory.  We will be able to see how the scalar is strongly coupled, and hence invisible when looking only at small fluctuations, but reappears when looking at higher-order fluctuations.  There is also a weakly coupled $D\rightarrow 4$ limit, achieved with a different scaling of the coupling, which gives $(\partial\phi)^4$ theory.  The $(\partial\phi)^4$ term has the correct positive sign dictated by analyticity constraints \cite{Adams:2006sv} if and only if the Gauss--Bonnet coupling $\alpha$ has the correct sign found in Ref.~\cite{Cheung:2016wjt}.  

 We will also argue that there can be no new purely gravitational four-point amplitudes beyond those given by the Lovelock theory \eqref{lovelocklag} by directly constructing on-shell four-point graviton amplitudes with up to six derivatives in arbitrary dimensions, showing that there is no purely gravitational ``novel 4D Gauss--Bonnet theory" from this point of view.  Lastly, we consider the cross section for unpolarized graviton-graviton scattering in GR and Gauss--Bonnet theory in arbitrary dimensions, showing that finite results from Gauss--Bonnet are possible for $D=4$ only if the coupling is scaled in yet another way.

\section{$D\rightarrow 4$ limits of Gauss--Bonnet amplitudes}

Let us attempt to find a new four-graviton amplitude by computing the amplitude in Gauss--Bonnet theory in $D$ dimensions using the Lagrangian \eqref{higherda} and then taking the $D\rightarrow 4$ limit.  To do this we must choose $D$-dimensional momenta and polarizations.  A natural choice is to choose those that lie only in four of the dimensions.  So for the momenta we will take
\be p^A=(p^\mu,0,\ldots,0),\ \ \ \ee
and for the polarizations we will take
\be \epsilon^{AB}=\left(\begin{array}{c|c}\epsilon^{\mu\nu} & 0 \\\hline 0 &  0  \end{array}\right),\ee
where $A,B,\dots$ are $D$-dimensional indices and $\mu,\nu,\dots$ are four-dimensional indices, $p^\mu$ are four-dimensional momenta, and $\epsilon^{\mu\nu}$ are four-dimensional graviton polarizations.  The amplitude does not trivially reduce to GR when $D=4$ because there are explicit factors of $D$ that appear apart from the polarizations and momenta. 
The result is the following non-vanishing helicity amplitudes:
\begin{subequations} \label{GBhelicityampse}
\begin{align} 
{\cal A}^{++,--} &=  {\cal A}^{--,++}=-{6\over M_P^{2+N}}\alpha^2 {N\over N+2}stu\, ,\\
{\cal A}^{++,++} &=  {\cal A}^{--,--}={1\over M_P^{2+N}} \left( {s^3\over tu}-2\alpha^2 {N\over N+2}s^3\right)\,,\\
{\cal A}^{+-,-+} &=  {\cal A}^{-+,+-}={1\over M_P^{2+N}} \left( {t^3\over su}-2\alpha^2 {N\over N+2}t^3\right)\, ,\\
{\cal A}^{+-,+-} &=  {\cal A}^{-+,-+}={1\over M_P^{2+N}} \left( {u^3\over st}-2\alpha^2 {N\over N+2}u^3\right)\, .
\end{align}
\end{subequations}
Here 
\be N \equiv D-4\ee
is the number of extra dimensions, the momenta are chosen so that the first two helicities are ingoing and the last two are outgoing, and the Mandelstam variables are defined in the usual way (we use the mostly plus metric),
\be s=-(p_1+p_2)^2,\ \ \ t=-(p_1-p_3)^2,\ \ \ u=-(p_1-p_4)^2\,.\label{mandelstamdefe}\ee
When $\alpha=N=0$, the amplitudes \eqref{GBhelicityampse} reduce to those of GR in four dimensions.  The absence of terms linear in $\alpha$ is consistent with the results of Ref.~\cite{Capper:1979pr}.

To get a finite correction as $N\rightarrow 0$ we make the replacement
\be \alpha \rightarrow {\tilde \alpha\over \sqrt{N}}\label{scaling2ee}\ee
and keep $\tilde\alpha$ finite as we take $N$ to zero.
This is different than the replacement $\alpha \rightarrow {\tilde \alpha/ {N}}$ of Eq.~\eqref{replaceorige} that is required for the classical solutions studied in Ref.~\cite{Glavan:2019inb}---if we were to attempt this scaling, our the amplitudes would blow up.  This already casts doubt on the universality of the prescription of Eq.~$\eqref{replaceorige}$ to give us all observables of a new theory; see also Refs.~\cite{Mahapatra:2020rds,Tian:2020nzb} for discussions about the non-uniqueness of the prescription.
We will see the explanation for this in the next section. 

\begin{subequations}\label{GBhelicityampse2}
After the replacement \eqref{scaling2ee}, the amplitudes at $N=0$ become
\begin{align} 
 {\cal A}^{++,--} &=  {\cal A}^{--,++}=-{3\over M_P^2}\tilde\alpha^2 stu\, ,\\
{\cal A}^{++,++} &=  {\cal A}^{--,--}={1\over M_P^2} \left( {s^3\over tu}-\tilde\alpha^2 s^3\right) \, ,\\
{\cal A}^{+-,-+} &=  {\cal A}^{-+,+-}={1\over M_P^2} \left( {t^3\over su}-\tilde\alpha^2 t^3\right)\, ,\\
{\cal A}^{+-,+-} &=  {\cal A}^{-+,-+}={1\over M_P^2} \left( {u^3\over st}-\tilde\alpha^2 u^3\right)\, . 
\end{align}
\end{subequations}
This is now an amplitude different from the GR amplitude.  But it is not a new purely gravitational amplitude.  Hidden inside it is a new scalar field.   The new $\tilde\alpha$-dependent terms can be accounted for by a massless scalar exchange pole.\footnote{The presence of a pole is clearer in spinor-helicity variables; for example, $\mathcal{A}^{++,--}\propto  \langle 1 2 \rangle^4 [34]^4(1/stu-\tilde{\alpha}^2/s)$ for all particles incoming.}  On the pole, the amplitude factors into a non-minimal four-derivative three-particle interaction between two gravitons and the scalar.  The amplitudes \eqref{GBhelicityampse2} are reproduced by the four-dimensional Lagrangian
\be S={M_P^2} \int d^4x\sqrt{-g} \bigg[{R\over 2}-{1\over 2}(\partial\phi)^2 +{1\over 2}\tilde\alpha \phi \, {\cal G} +\dots \bigg]\, ,\label{firstgbampee}
\ee
where $\phi$ is the new scalar. Note that this scalar-tensor theory differs from the one described in Refs.~\cite{Lu:2020iav, Kobayashi:2020wqy}. We explain the relation between them in the next section.

Given that there is a scalar degree of freedom, we should also be able to compute amplitudes involving external scalars.  We can do this by using the following $D$-dimensional polarization: for a graviton moving in the $\hat {\bf z}$ direction, we take
\be  \epsilon^{AB}(\hat {\bf z})= {1\over \sqrt{{1\over 2}+{1\over N}}}\left(\begin{array}{c|c}  {1\over 2}\left(\begin{array}{cccc}0 & 0 & 0 & 0 \\0 & 1 & 0 & 0 \\0 & 0 & 1 & 0 \\0 & 0 & 0 & 0\end{array}\right)  & 0 \\\hline 0 & -{1\over N}\delta_{mn}   \end{array}\right)\,, 
\label{scalarpole}
\ee
where $n,m=1,\ldots,N$ are the extra-dimensional components.  
This is properly normalized and traceless in $D$ dimensions, and transverse to the momentum traveling in the $\hat{\bf z}$ direction, $p^A=(E,0,0,E,0,\dots,0)$.

Using the polarizations defined by Eq.~\eqref{scalarpole} for all external legs in the $D$-dimensional four-point Gauss--Bonnet amplitude, we obtain
\be   {\cal A}= {\left(s^2+t^2+u^2\right)^2\over 4  M_P^{2+N} stu}+  {4\left(s^2+t^2+u^2\right)\over  M_P^{2+N} N(N+2)}\alpha-\frac{6 \left(N^3-2 N^2-12 N+8\right)stu}{ M_P^{2+N} N (N+2)^2}\alpha^2\,.\label{scalar4phiampe}
\ee
This amplitude now blows up as $1/N$ as $N\rightarrow 0$.  To obtain a finite correction as $N\rightarrow 0$, we make the replacement
\be \alpha \rightarrow {N \tilde \alpha}\label{scaling3ee}\ee 
and hold $\tilde{\alpha}$ fixed as $N$ goes to zero. This gives
\be {\cal A}= {1\over 4M_P^2}{\left(s^2+t^2+u^2\right)^2\over stu}+{2\over M_P^2}\tilde \alpha \left(s^2+t^2+u^2\right)\,. \label{phi4ampe}\ee
This is now a four-scalar amplitude.  In fact, it is the amplitude obtained from the minimally coupled $(\partial\phi)^4$ theory,
\be S=M_P^2\int d^4x\sqrt{-g} \bigg[{1\over 2}R  -{1\over 2}(\partial{\hat\phi})^2 +\tilde\alpha (\partial{\hat\phi})^4   \bigg].\label{phil4theorylage}\ee
The $\tilde\alpha$-dependent term in \eqref{phi4ampe} comes from the $(\partial\phi)^4$ contact interaction and the rest comes from graviton exchange between the scalars.

\section{Lagrangians}

We can understand the origin of the above results from the dimensional reduction of the Lagrangian \eqref{higherda}.  Following Ref.~\cite{Lu:2020iav},\footnote{We thank Tony Padilla for sharing some notes last summer that contained essentially the same results.} consider dimensionally reducing on a flat $N$-dimensional manifold. The reduction ansatz for the  metric is
\be 
ds_D^2= g_{\mu\nu}(x)dx^\mu dx^\nu+e^{2\phi(x)}d\vec{y}_N^{\ 2},\label{dimredans}
\ee
where $D=4+N$, $d\vec{y}_N^{\ 2}$ is the flat metric for the extra dimensions, $g_{\mu\nu}$ is the four-dimensional metric and $\phi$ is a scalar corresponding to the volume modulus (the dilaton).  All the other Kaluza--Klein fields are being set to zero.\footnote{This is a consistent truncation, which can be seen as follows.  The metric and dilaton being kept in the ansatz of Eq.~\eqref{dimredans} are the only fields whose wave functions in the extra dimensions are constant scalar zero modes \cite{Hinterbichler:2013kwa}.  Consider possible cubic terms that contain a single instance of one of the missing Kaluza--Klein fields.  These would be terms where a removed Kaluza--Klein mode sources the modes being kept, potentially spoiling the consistency of the truncation.  All such terms vanish since they involve a triple overlap integral of two zero modes, which are constant and can be pulled out of the integral, with a non-zero mode which always integrates to zero or comes with a derivative that kills one of the zero modes \cite{Bonifacio:2019ioc}.}
Integrating over the extra dimensions gives \cite{Lu:2020iav, Charmousis:2012dw, Charmousis:2014mia}
\begin{align} S=&{M_4^2\over 2}\int d^4x\sqrt{-g} e^{N \phi} \bigg[R  +\alpha\,{\cal G} + N(N-1) (\partial\phi)^2 \nn\\
&- \alpha N(N-1) \left(4G^{\mu\nu}\partial_\mu\phi\partial_\nu\phi +2 (N-2)\square\phi (\partial\phi)^2+(N-1)(N-2)(\partial\phi)^4   \right)\bigg], \label{actiondimeredbde}
\end{align}
where $M_4^2={\cal V}_N M_P^{D-2}$ with ${\cal V}_N$ the volume of the extra dimensions and $G_{\mu\nu}$ is the Einstein tensor of $g_{\mu\nu}$.

We can now consider this as a four-dimensional Lagrangian depending on some parameter $N$.  
If we just set $N=0$ in \eqref{actiondimeredbde}, we get back Einstein gravity (as expected since we are not reducing any dimensions).  In this case we lose the scalar degree of freedom.
If we instead replace $\alpha\rightarrow {\tilde\alpha/ N}$, as in Eq.~\eqref{replaceorige}, and then set $N=0$ while holding $\tilde{\alpha}$ fixed, we get\footnote{This Lagrangian can also be obtained by the limiting procedure of Ref.~\cite{Mann:1992ar} applied to Gauss--Bonnet theory \cite{Fernandes:2020nbq, Hennigar:2020lsl}.} (taking ${\cal V}_0=1$)
\begin{equation}  \label{eq:StronglyCoupled}
S={M_P^2\over 2}\int d^4x\sqrt{-g} \bigg[R    
+ \tilde \alpha  \left(\phi \,{\cal G}+4G^{\mu\nu}\partial_\mu\phi\partial_\nu\phi -4\square\phi (\partial\phi)^2+2(\partial\phi)^4   \right)\bigg].
\end{equation}
In this case we have not lost any degrees of freedom: the scalar is still there and there are no extra degrees of freedom beyond the scalar since this theory belongs to the ghost-free Horndeski class of scalar-tensor theories \cite{Horndeski:1974wa}.

The Lagrangian in Eq.~\eqref{eq:StronglyCoupled} does not have a kinetic term for $\phi$.  This means that $\phi$ is infinitely strongly coupled around the flat background.  
We can see this more explicitly by first diagonalizing the action \eqref{actiondimeredbde} by making a conformal transformation $g_{\mu\nu}\rightarrow e^{-N\phi}g_{\mu\nu}$.  The action then becomes, up to total derivatives,
\begin{align} 
S&=M_4^2\int d^4x\sqrt{-g} \bigg[{1\over 2}R  -{1\over 4}N(N+2)(\partial\phi)^2  +{1\over 2}\alpha e^{N\phi}{\cal G}  \nn\\
&+{1\over 2} \alpha N e^{N\phi} \left(4G^{\mu\nu}\partial_\mu\phi\partial_\nu\phi + (N-2)(N+2)\square\phi (\partial\phi)^2-(N-1)(N+2)(\partial\phi)^4   \right)\bigg]. \label{actiondimeredbde2}
\end{align}
Now the kinetic terms have been diagonalized and we see that the kinetic term for $\phi$ vanishes if we set $N=0$.  To preserve the scalar kinetic term we should first canonically normalize by replacing $\phi\rightarrow {\hat \phi/ \sqrt{N}}$, 
\begin{align} S&=M_4^2\int d^4x\sqrt{-g} \bigg[{1\over 2}R  -{1\over 4}(N+2)(\partial{\hat\phi})^2  +{1\over 2}\alpha e^{\sqrt{N}{\hat\phi}}{\cal G}  \nn\\
&+{1\over 2} \alpha  e^{\sqrt{N}{\hat\phi}} \left(4G^{\mu\nu}\partial_\mu{\hat\phi}\partial_\nu{\hat\phi} + {(N-2)(N+2)\over \sqrt{N}}\square{\hat\phi} (\partial{\hat\phi})^2-{(N-1)(N+2)\over N}(\partial{\hat\phi})^4   \right)\bigg]. \label{actiondimeredbde22}
\end{align}
Now we see the cubic galileon $\square{\hat \phi} (\partial{\hat \phi})^2$ and the $(\partial \hat \phi)^4$ self-couplings going to infinity as $N\rightarrow 0$.  

We thus see that the flat space solution is infinitely strongly coupled with the scaling \eqref{replaceorige} used in Ref.~\cite{Glavan:2019inb} and in many of the follow-up studies.  Thus it is not really a new gravity theory, it is a scalar-tensor theory where the scalar happens to be invisible in quadratic fluctuations because its kinetic term vanishes.  The scalar would reappear when higher-order couplings are considered, as they are in the scattering amplitudes.  We believe it is likely that something similar is happening around the other highly symmetric solutions being studied, such as black holes and FLRW solutions.

To keep the scalar self-interactions in Eq.~\eqref{actiondimeredbde22} finite, we should instead be making the replacement $\alpha\rightarrow {N\tilde\alpha}$ with $\tilde{\alpha}$ fixed.  When $N=0$ we then get
\be S=M_P^2\int d^4x\sqrt{-g} \bigg[{1\over 2}R  -{1\over 2}(\partial{\hat\phi})^2 +\tilde\alpha (\partial{\hat\phi})^4   \bigg].\label{phi4lag332e}\ee
This is just the $(\partial\hat{\phi})^4$ theory \eqref{phil4theorylage} with the correct coefficients to produce the amplitude we saw in \eqref{phi4ampe}, with precisely the scaling \eqref{scaling3ee}.  Note that the polarization \eqref{scalarpole} used to compute the amplitude with external scalars corresponds to a fluctuation that changes the overall volume but not the shape of the extra dimensions, which is precisely the dilaton mode.
Additionally, notice that the  $(\partial\hat{\phi})^4$ term in Eq.~\eqref{phi4lag332e} has the correct sign as dictated by analyticity constraints \cite{Adams:2006sv} (which have also been shown to hold in certain curved spaces \cite{Hartman:2015lfa}) if and only if the Gauss--Bonnet coupling $\alpha$ has the correct sign \cite{Cheung:2016wjt} (note however that there are subtleties due to the presence of the graviton pole \cite{Bellazzini:2019xts}).

If we scale $\alpha\rightarrow\sqrt{N}\tilde\alpha$ and ignore the singular $(\partial \hat \phi)^4$ term, then the galileon interaction $-2M_P^2\tilde\alpha\, \square{\hat\phi} (\partial{\hat\phi})^2$ is the next subleading self-interaction in Eq.~\eqref{actiondimeredbde22} as $N\rightarrow 0$.  The exchange amplitude produced from this cubic interaction is ${\cal A}=-12\tilde\alpha^2stu/M_P^2$.  This matches the last term in Eq.~\eqref{scalar4phiampe} when $N=0$. 

Suppose that we ignore all the scalar self-interactions and consider only the terms in Eq.~\eqref{actiondimeredbde22} that can contribute to four-graviton scattering.  The only important term is then where a single $\phi$ enters, which can serve as a coupling mediating $\phi$ exchange,
\be  
S=M_4^2\int d^4x\sqrt{-g} \bigg[{1\over 2}R  -{1\over 4}(N+2)(\partial{\hat\phi})^2  +{1\over 2}\alpha{\sqrt{N}{\hat\phi}}\,{\cal G}+\cdots\bigg]\,.  \label{actiondimeredbde223}
\ee
To get a nontrivial result when $N=0$ we now have to scale $\alpha \rightarrow {\tilde \alpha/ \sqrt{N}}$.
This is the same Lagrangian as in Eq.~\eqref{firstgbampee}, which has the correct terms to produce the amplitudes \eqref{GBhelicityampse2} with precisely the scaling \eqref{scaling2ee}. However, with this scaling the scalar self-interactions in Eq.~\eqref{actiondimeredbde22} blow up, as would other amplitudes involving too many external scalars.   

To summarize, the only true nontrivial $N\rightarrow 0$ limit which is not infinitely strongly coupled is the one where we scale $\alpha\rightarrow {N\tilde\alpha}$, which leads to Eq.~\eqref{phi4lag332e}. Therefore, in this sense, the only nontrivial $D=4$ limit of Gauss--Bonnet without infinite strong coupling around flat space is minimally coupled $(\partial\phi)^4$ theory.

\section{No new graviton amplitudes}

We have seen that the dimensional continuation and $D\rightarrow 4$ procedure applied to Gauss--Bonnet amplitudes does not produce new pure gravity amplitudes, but rather reproduces amplitudes from known scalar-tensor theories.  One could argue that perhaps the procedure we used, of simply taking momenta and polarizations to lie in the four-dimensional space, is not general enough and that some other more complicated procedure produces the new $D=4$ pure gravity amplitudes.  Here we apply the on-shell construction of graviton amplitudes in general dimensions to argue that this cannot be, since there are no four-point amplitudes with no more than six derivatives and the usual properties of Lorentz invariance, unitarity and locality other than those of Lovelock gravity.

Consider four-point pure gravity tree amplitudes.  By factorization, such amplitudes must be of the form
\be {\cal A}={\cal A}_{\rm exchange}+{\cal A}_{\rm contact},\ee
where ${\cal A}_{\rm contact}$ is analytic in the momenta and ${\cal A}_{\rm exchange}$ has simple poles with residues that factorize into products of on-shell three-point graviton amplitudes.
Gauge invariance tells us that the full amplitude must satisfy the Ward identity on all external legs, $\delta{\cal A}=0$.  In general, ${\cal A}_{\rm exchange}$ is not gauge invariant by itself, but rather its gauge variation is an analytic term that is cancelled by the gauge variation of ${\cal A}_{\rm contact}$.

Consider first amplitudes that involve at most two powers of the momenta. The three-point amplitudes are fixed by Lorentz invariance and are fully classified.  There are a finite number of possible three-point functions for each collection of three spins and masses.  The only gauge-invariant on-shell three-point amplitude that involves three massless spin-2 particles and two powers of momenta is proportional to the Einstein--Hilbert amplitude,
\be {\cal A}_{3,{\rm EH}} \propto  \left[ (p_1\cdot  \epsilon_3)  (\epsilon_1\cdot  \epsilon_2)+ (p_3\cdot  \epsilon_2)  (\epsilon_1\cdot  \epsilon_3) + (p_2\cdot  \epsilon_1)  (\epsilon_2\cdot  \epsilon_3) \right]^2\,,\label{EHvertexe}\ee
where the polarization tensors are written as $\epsilon_{\mu \nu} = \epsilon_{\mu} \epsilon_{\nu}$.
We form ${\cal A}_{\rm exchange}$ out of this vertex by gluing together two copies with a propagator and summing over all three channels.  The gauge variation $\delta{\cal A}_{\rm exchange}$ is a polynomial with two powers of momenta, which must be cancelled by the variation of some ${\cal A}_{\rm contact}$ with two derivatives.  The form of ${\cal A}_{\rm contact}$ is uniquely fixed by requiring $\delta{\cal A}_{\rm contact}=-\delta{\cal A}_{\rm exchange}$ and the resulting amplitude matches the GR amplitude, showing that it is the unique two-derivative amplitude in any number of dimensions.

The Gauss--Bonnet term has four derivatives, so to study the possible amplitudes that could result from some kind of dimensional continuation of Gauss--Bonnet, we need to study amplitudes with more than two powers of the momenta.   A four-derivative term could give rise to three-point functions with up to four powers of momenta and, through exchange diagrams, four-point functions with up to six powers of momenta.  There are two gauge invariant on-shell three-point amplitudes that involve the massless spin-2 particles and up to four powers of momenta, the Einstein Hilbert vertex in Eq.~\eqref{EHvertexe} and the Gauss--Bonnet vertex,
\be 
{\cal A}_{3,{\rm GB}} \propto (p_1\cdot  \epsilon_3)(p_2\cdot  \epsilon_1)(p_3\cdot  \epsilon_2 )\Big[(p_1\cdot \epsilon_3)(
\epsilon_1\cdot  \epsilon_2) + (p_3\cdot  \epsilon_2)(\epsilon_1\cdot \epsilon_3) + 
(p_2\cdot  \epsilon_1)(\epsilon_2\cdot \epsilon_3)\Big]\,.\label{GBvertexe}
\ee 
Taking as the cubic vertex a general linear combination of these two and forming the exchange diagrams, we get an ${\cal A}_{\rm exchange}$ with up to six powers of momenta.  We thus write an ansatz for ${\cal A}_{\rm contact}$ with up to six momenta.  The parameters of this ansatz are not completely fixed by requiring $\delta{\cal A}_{\rm contact}=-\delta{\cal A}_{\rm exchange}$, rather there is one free parameter left over.   If the coefficient of the two-derivative Einstein-Hilbert cubic is non-zero, the resulting amplitude is precisely what one obtains from Lovelock theory \eqref{lovelocklag}.  The extra parameter in the contact ansatz corresponds to the quartic part of the $R^3$ Lovelock interaction (these terms do not contribute a cubic vertex, due to the Lovelock structure).  The Lovelock terms of order $R^4$ and higher do not contribute to the four-point amplitude.  When $D=4$, this amplitude reduces to the GR amplitude.  If the coefficient of the two-derivative Einstein-Hilbert cubic is zero, the resulting amplitude is that of the pseudolinear Lovelock theory discussed in Refs.~\cite{Hinterbichler:2013eza,Chatzistavrakidis:2016dnj,Bai:2017dwf,Bonifacio:2019pfg}, which vanishes in $D=4$.

This argument shows quite generally that there is no new purely gravitational amplitude that could possibly arise from any kind of $D\rightarrow 4$ limit of the Gauss--Bonnet action, assuming that the limit does not involve somehow adding more derivatives and does not spoil Lorentz invariance, unitarity, or locality.

\section{Summing over spins}

Another possible way to obtain four-dimensional observables from Gauss--Bonnet theory is to consider the unpolarized differential cross section in $D$ dimensions and to take $D=4$ after rescaling parameters.  In $D$ dimensions the unpolarized differential cross section for $2\rightarrow 2$ elastic scattering of identical particles in the center-of-mass frame is given by  
\be 
\frac{d \sigma}{d \Omega_{D-2}}= {1\over 2}{1\over (2\pi)^{D-2}}{E^{D-6}\over 64} {1\over {{D(D-3)\over 2 }}} \sum_{\rm helicities} \left|{\cal A}\right|^2.
\ee
Here $E$ is the energy of one of the particles, the sum runs over the helicities of each particle, the factor of $1/2$ is a symmetry factor because the final state particles are identical, and the denominator $D(D-3)/2$ is the  number of graviton states in $D$ dimensions and is there because we average over initial states and sum over final states.

For the Gauss--Bonnet theory \eqref{higherda}, performing the sum in general dimensions is computationally intensive,\footnote{To do this we use the polarization sum \cite{vanDam:1970vg}
\be \sum_\lambda  \epsilon^{A B}_{(\lambda)}(p)\epsilon^{\ast C D}_{(\lambda)}(p)={1\over 2}(\Pi^{A C }\Pi^{B D}+\Pi^{A D}\Pi^{B C })-\frac{1}{D-2}\Pi^{A B }\Pi^{C D}\ee
for each of the particles in the squared amplitude.   Here $\Pi^{A B }\equiv\eta^{A B }+\left(p^A \bar p^B +p^B \bar p^A \right)$, where $p^A =(E,\pb)$ is the on-shell $D$-momentum and $\bar p^A $ is the unique vector satisfying $\bar p^A \bar p_A =0$, $p^A \bar p_A =-1$, given by $ \bar p^A ={1\over 2E^2}\left(E,-\pb\right).$} since we cannot just calculate each individual polarization and sum the results like you can in a fixed dimension or use spinor-helicity variables as in $D=4$.  The result is
\begin{align} 
M_P^{2(D-2)} \sum_{\rm helicities}   \left|{\cal A}\right|^2&=
   \frac{(D-3) D^2}{4 {S_3}^2} \left((D-3) {S_2}^4+2 (D-6) {S_2} {S_3}^2\right)  \nn\\
   & +\frac{(D-4) (D-3) D}{2 (D-2)} \bigg[-\alpha    \left(2(D-3) D  \frac{{S_2}^3 }{S_3}+3 (D-6) (D-2)
   {S_3}\right)  \nn\\
&+\alpha ^2\frac{ \left(9 D^3-52 D^2+76 D-24\right)
   }{D-2}{S_2}^2  -\alpha ^3\frac{ \left(D^2-8
   D+36\right) \left(7 D^2-20 D+4\right) }{(D-2)^2}{S_2} {S_3}  \nn\\
&+\alpha ^4\frac{1}{(D-2)^3 D} \bigg( 6 \left(5 D^6-61 D^5+292 D^4-784 D^3+1600 D^2-1840 D+896\right) {S_3}^2   \nn\\
&-(D-2)^2 D (D+2) \left(D^2-4
   D+12\right) {S_2}^3 \bigg) \bigg]\,,
\label{gaussbonnetspins}
\end{align}
where $S_2$ and $S_3$ are the two independent on-shell symmetric polynomials of the Mandelstam variables \eqref{mandelstamdefe},
\be S_2\equiv st+su+tu,\quad S_3 \equiv stu.\ee

When $\alpha=0$, this reduces to the GR result, which can be written as
\be \sum_{\rm helicities} \left|{\cal A}\right|^2=\frac{(D-3) D^2}{16 M_P^{2(D-2)} s^2 t^2 u^2}  \left[(6-D) \left(s^8+t^8+u^8\right)+6 (D-4) (s t+s u+t u)^4\right]\,.\ee
This result is well known in $D=4$ \cite{DeWitt:1967uc,Berends:1974gk,Grisaru:1975bx}, but as far as we are aware the result for general $D$ has not appeared before and neither has the Gauss--Bonnet result of Eq.~\eqref{gaussbonnetspins}.

In Eq.~\eqref{gaussbonnetspins} all the polynomials in $D$ have been factored over the integers as much as possible.  There is a single factor of $D-4$ in front of the $\alpha$-dependent terms, so they vanish in $D=4$ as expected.  Since the highest power of $\alpha$ is $\alpha^4$, to get a nontrivial correction we must take yet another different scaling where we make the replacement 
\be \label{eq:1/4rescale}
\alpha \rightarrow {\tilde\alpha\over (D-4)^{1/4}}
\ee
and hold $\tilde{\alpha}$ fixed.
After this rescaling, Eq.~\eqref{gaussbonnetspins} simplifies in $D=4$ to 
\be M_P^{4} \sum_{\rm helicities}\left|{\cal A}\right|^2=4\left({S_2^4\over S_3^2}-4S_2\right)+36\tilde\alpha^4 \left(9S_3^2-S_2^3\right).\label{new4dcsecee}\ee
The sum over helicities involves many of the Kaluza--Klein modes that were discarded in the ansatz in Eq.~\eqref{dimredans}.  Reproducing this and the scaling \eqref{eq:1/4rescale} from a four-dimensional theory would thus presumably involve keeping track of these extra modes and summing over their amplitudes. The fact that GR gives the only four-point amplitude with at most six powers of momenta in $D=4$ means that Eq.~\eqref{new4dcsecee} cannot come from spin averaging any four-dimensional pure gravity amplitude.

\section{Conclusions}

The question of whether there exists a ``novel 4D Gauss--Bonnet theory" has been revived in Ref.~\cite{Glavan:2019inb}.   Observables in such a theory are supposed to be obtained by taking a $D\rightarrow 4$ limit of Gauss--Bonnet observables in general $D$ after rescaling the Gauss--Bonnet coupling by $\alpha\rightarrow {\tilde \alpha/(D-4)}$, leaving a finite contribution which differs from GR.  We have studied this from the point of view of scattering amplitudes, which are natural observables of any theory in flat space and should presumably be calculable with such a prescription, even if there is no strict four-dimensional Lagrangian or equations of motion.  

The most natural way of taking the four-dimensional limit of Gauss--Bonnet amplitudes leads to the amplitudes of a scalar-tensor theory.  This is consistent with the conclusions of Refs.~\cite{Lu:2020iav,Fernandes:2020nbq,Hennigar:2020lsl} (see also \cite{Arrechea:2020evj,Gabadadze:2020tvt,Aoki:2020lig,Shu:2020cjw,Samart:2020sxj}).  The scalar in this theory is infinitely strongly coupled around flat space when using the usual scaling $\alpha\rightarrow {\tilde \alpha/(D-4)}$. This suggests that the same is true around other highly symmetric solutions and would explain why it is not seen in small fluctuations about these solutions.  To keep the scalar self-couplings finite, a scaling $\alpha\rightarrow (D-4){\tilde \alpha}$ should be used, which leads to a $(\partial\phi)^4$ theory, with the correct sign as dictated by analyticity.

The fact that solutions in the novel Gauss--Bonnet theory differ from GR could be explained by the fact that they are sourced by this extra scalar.  The fact that the scalar is infinitely strongly coupled means that it only shows up in higher-order fluctuations, which the scattering amplitudes are sensitive to.  It would be interesting to explore higher-order fluctuations around some of the other solutions, such as black holes or FLRW backgrounds, to see if the scalar is visible there too.  

We also looked at the unpolarized graviton-graviton cross section to see that another distinct rescaling of the coupling is needed to obtain non-trivial finite corrections in four dimensions.

Finally, we argued that there are no new purely gravitational four-point tree amplitudes that could result from a $D\rightarrow 4$ limit of Gauss--Bonnet by directly constructing the on-shell amplitudes, and so in this sense there is no ``novel 4D Gauss--Bonnet theory."

{\bf Acknowledgments:} We would like to thank Nathan Moynihan, Tony Padilla, Glenn Starkman, and Bayram Tekin for helpful conversations and correspondence, and Tony Padilla for sharing some of his notes.  KH and JB acknowledge support from DOE grant DE- SC0019143 and Simons Foundation Award Number 658908.

\bibliographystyle{utphys}
\addcontentsline{toc}{section}{References}
\bibliography{GB-arxiv-v2}

\end{document}